\documentstyle[11pt]{article}
\input psfig.tex
\begin{document}
\centerline{\bf\large
A Review of pQCD Calculations of Electromagnetic }
\centerline{\bf \large Form Factors of
Hadrons
}

\bigskip

\centerline{\bf Pankaj Jain$^a$, Bijoy Kundu$^a$, Hsiang-nan Li$^b$,}
\centerline{\bf 
John P. Ralston$^c$ and
Jim Samuelsson$^d$}

\bigskip 
\centerline{$^a$Department of Physics, IIT Kanpur, Kanpur-208 016, India
 }

\centerline{$^b$Department of Physics, National Cheng-Kung University,
Tainan,}
\centerline{ Taiwan, Republic of China}
 \centerline{$^c$Department of Physics \& Astronomy, University of
Kansas,}
\centerline{
Lawrence, KS 66045, USA}
       \centerline{$^d$Department of Theoretical Physics,
 Lund University, Lund, Sweden
}

\begin{abstract}
We review the current status of perturbative QCD calculation of
hadronic electromagnetic form factors.
\end{abstract}

\section{INTRODUCTION}

The applicability of perturbative QCD to exclusive processes at large
momenta is an
interesting research problem. The Brodsky-LePage \cite{BL} pQCD based
factorization
has been only partially successful.
In this case the process is factorized into a perturbatively calculable
hard scattering piece and the soft distribution amplitude.
The pion electromagnetic
form factor \cite{BL,FJ,ER} at momentum transfer $q^2 = -Q^2$, for example,
can be written as
\begin{equation}
F_\pi(Q^2) = \int dx_1 dx_2 \phi(x_2,Q) H(x_1,x_2,Q) \phi(x_1,Q)
\end{equation}
where $\phi(x,Q) $ are the distribution amplitudes which can
be expressed in terms of the pion wave function $\psi(x,\vec k_T)$ as
\begin{equation}
 \phi(x,Q) = \int^Q d^2k_T\psi(x,\vec k_T) .
\end{equation}
Here $x$ is the longitudinal momentum fraction and $\vec k_T$ the
transverse momentum carried by the quark. The factorization is
possible provided the external photon momentum $Q^2$ is much larger
than the intrinsic quark transverse momentum $k_T^2$, in which case
the $k_T$ dependence of the hard scattering $H$ can be neglected.

The formalism predicts that at large momenta the cross section for
exclusive processes
$d\sigma/dt$, where $t$ is the momentum transfer squared, scales
like $1/t^{n-2}$ up to logs, where $n$ is the total number of elementary
partons participating in the process. The underlying reason
for the power law is
scale invariance of the fundamental
theory. The extra logarithmic
dependence is given by QCD scaling violations.
The dominant contribution to this scattering arises from the valence
quark, since every additional parton leads to an additional suppression
factor of $1/t$. Physically the scattering probes the short distance
part of the hadron wave function. Dominance by the short distance
wave functions leads to several predictions such as helicity conservation,
color transparency  \cite{BM,JPR} etc.

The successes and failures of this scheme
are well known. The predicted momentum dependence of exclusive
processes, in particular the hadronic electromagnetic form factors,
 have generally been found to be in good agreement with data.
However more detailed dynamical predictions such as helicity conservation
in hadron-hadron collisions fail to agree. Calculation
of electromagnetic form factors using this factorization
scheme has been criticised by several authors. The basic problem
is that the momentum scales of the exchanged gluons tend to become
rather small, and the applicability of pQCD becomes doubtful. The
normalization of form factors is largely unknown; use of asymptotic
distribution
amplitudes tends to give small normalizations compared to data. Form
factor magnitudes can
be enhanced by use of model distribution amplitudes which peak closer to the
endpoints, namely $x\rightarrow 0,1$, which then exacerbates the problem
of small internal momentum transfers.

\section{THE SUDAKOV FORM FACTOR}

In order to investigate this problem
in more detail, Botts and Sterman \cite{BS}
and Li and Sterman \cite{LS} developed an alternate factorization which
does not neglect the $k_T$ dependence of the hard scattering. This
formalism also
includes use of a Sudakov form factor. For the case of pion form factor
\cite{LS}
the starting point is,
\begin{equation}
F_\pi(Q^2) = \int dx_1 dx_2 d\vec k_{T1} d\vec k_{T2} \psi^*(x_2,\vec
k_{T2},P_2)
H(x_1,x_2,Q^2,\vec k_{T1},\vec k_{T2}) \psi(x_1,\vec k_{T1},P_1),
\end{equation}
where it is again assumed that the process is factorizable into
hard scattering and soft hadronic wave functions $\psi(x,\vec k_T,P)$. The
calculation is simplified by dropping the $k_T$ dependence in the
quark propagators in hard scattering kernel $H$,
in which case only the combination $\vec k_{T1} + \vec k_{T2}$
appears in the calculation.
The authors \cite{LS} work in configuration space where this can be written
as
\begin{equation}
F_\pi(Q^2) = \int dx_1 dx_2 {d^2\vec b\over (2\pi)^2}  {\cal P}(x_2,b,P_2,\mu)
\tilde H(x_1,x_2,Q^2,\vec b,\mu) {\cal P}(x_1,b,P_1,\mu),
\end{equation}
where ${\cal P}(x,b,P,\mu)$ and $\tilde H(x_1,x_2,Q^2,\vec b,\mu)$ are the
Fourier transforms of the wave function and hard scattering
respectively;
$\vec b$ is conjugate to $\vec k_{T1} + \vec k_{T2}$, $\mu$ is the
renormalization
scale and $P_1$, $P_2$ are the initial and final momenta of the pion.

Sudakov form factors are obtained
by summing the leading and next
to leading logarithms using renormalization group (RG) techniques.
The wave function at small $b$ with a transverse momentum $k_T$ cutoff
equal to $\omega=1/b$ can be approximated by the distribution amplitude
$\phi(x,1/b)$.
Large $k_T$ corrections can be evaluated perturbatively, which result in the
Sudakov form factor. The final result is given by:
\begin{eqnarray}
{\cal P}(x,b,P,\mu) &=& \exp\left[-s(x,\omega,Q)  - s(1-x,\omega,Q) -
2\int^\mu_\omega
{d\bar \mu\over \bar \mu} \gamma_q(\alpha_s(\bar\mu))\right] \nonumber\\
&\times &
\phi(x,1/b) + O(\alpha_s(\omega)).
\end{eqnarray}
where $\gamma_q(\alpha_s)$ is the quark anomalous dimension.
The explicit formula for the function $s(x,\omega,\mu)$ is given
in Li and Sterman \cite{LS}.
Here $\omega=1/b$ plays the role of the factorization scale,
above which QCD corrections give the perturbative evolution of the wave
function $P$, and below which QCD corrections are
absorbed into the nonperturbative distribution amplitude $\phi$. For the
case of the pion,
$1/b$ is the natural choice for this scale. However as discussed below,
for the proton the relevant scale is not obvious and several possibilities
exist in the literature.

\begin{figure} [t,b]
\hbox{\hspace{6em}
 \hbox{\psfig{figure=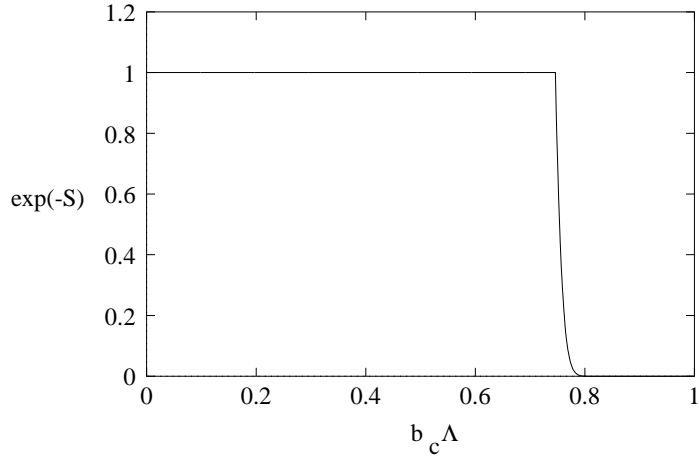,height=6cm}}}
  \caption {\em The Sudakov form factor $\exp(-S)$ with $Q^2 = 4 $ GeV$^2$.
For this calculation the QCD scale parameter $\Lambda$ was taken to
be $0.1$ GeV.
    }
  \label{pion2}
\end{figure}

The final formula for the form factor, after incorporating the
renormalization group evolution of the hard scattering from the
renormalization scale
$\mu$ to $t$, $t={\rm max}(\sqrt{x_1x_2}Q,1/b)$, is given by\cite{LS},

\begin{eqnarray}
F_\pi(Q^2) &=& 16\pi C_F \int^1_0 dx_1 dx_2 \phi(x_1)\phi(x_2)\int_0^\infty
b db\alpha_s(t) K_0(\sqrt{x_1x_2}Qb)\nonumber \\
&\times & \exp[-S(x_1,x_2,\omega,Q)],
\end{eqnarray}

where $$S(x_1,x_2,b,Q) = \sum_{i=1}^2\left[s(x_i,b,Q) + s(1-x_i,b,Q)\right]
-4\int_\omega^t{d\bar \mu\over\bar\mu}\gamma_q(\alpha_s(\bar\mu)).$$
The function $e^{-S}$ is plotted in fig. \ref{pion2}. It cuts off large
$b$ regions of the integral and hence the calculation is infrared finite,
without
needing any arbitrary infrared cutoff such as a gluon mass. At
small $b$ the function has been set equal to one.

\begin{figure} [t,b]
\hbox{\hspace{6em}
 \hbox{\psfig{figure=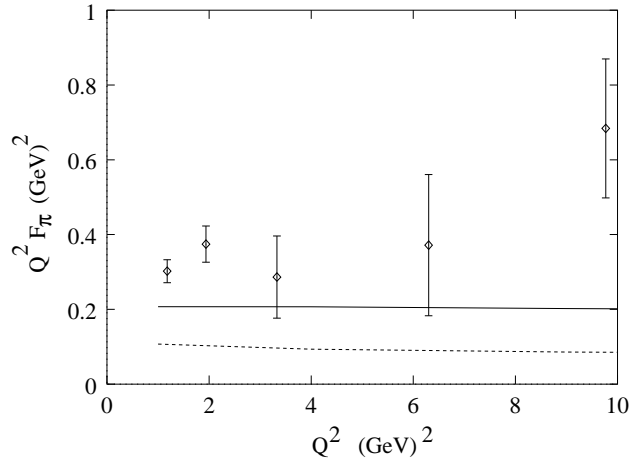,height=6cm}}}
  \caption {\em The pion form factor $F_\pi (Q^2)$
using the asymptotic (dotted line) and the CZ
    (solid line) distribution amplitudes.
    The experimental data, taken from Ref. 9 with errorbars are
also shown.}
  \label{pion1}
\end{figure}

The resulting form factor using asymptotic as well as  CZ \cite{CZ}
distribution amplitudes is shown in fig. \ref{pion1}. A remarkable
fact is that the correct asymptotic $Q^2$ behavior is seen beyond the scale
of about $Q=1$ GeV,
irrespective of the choice of wave function. In contrast to the Brodsky-LePage
factorization, the $k_T$ dependence of the hard scattering is not
neglected, and
hence this $Q^2$ dependence
does not follow trivially.  It is rather
a detailed dynamical prediction of the theory and depends on the relative
size of intrinsic $k_T^2$ and $x_1 x_2 Q^2$. The prediction is robust, since
it is independent of the details of the distribution amplitude used.
This simple yet important point
justifies the basic idea of Brodsky-LePage factorization, namely that $k_T$
can be treated as negligible in the hard scattering.

\begin{figure} [t,b]
\hbox{\hspace{6em}
 \hbox{\psfig{figure=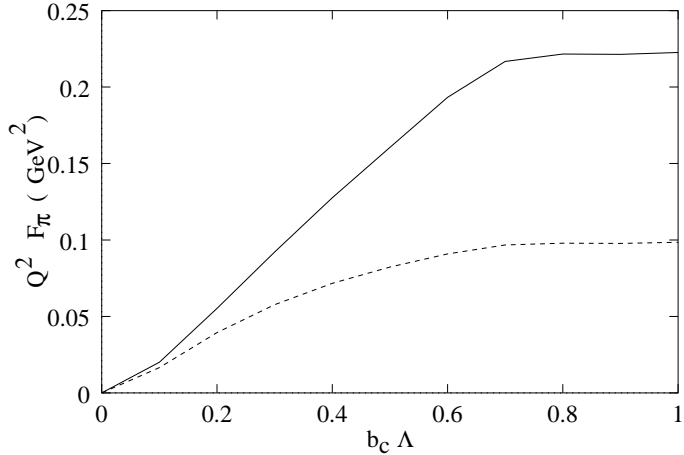,height=6cm}}}
  \caption {\em Dependence of $Q^{2}F_{\pi}(Q^2)$ on the transverse
distance cutoff $b_{c}$
    with the asymptotic (dotted line),
    and CZ (solid line) distribution amplitudes for $Q^2=4$ GeV$^2$.
	The QCD scale parameter $\Lambda$ was chosen to be $0.1$ GeV for this
calculation.}
  \label{pion3}
\end{figure}

We note that the normalization of the theoretical result falls below the
experimental data for both choices of distribution amplitude.
However, the large difference between theory and experiment at high momenta
should be interpreted with caution, since as emphasized by Sterman
and Stoler \cite{SS}, there may be large systematic errors in the experimental
extraction of the form factor which are not shown in the figure.
Further theoretical issues in this extraction have been raised in
Ref. \cite{CM}.

In any event, the theoretical normalization of the form factor is comparatively
murky, because it
depends on the details of the distribution amplitude.
Furthermore, the leading order pQCD amplitudes that are practical to
calculate may not give a very
reliable estimate of the
normalization. One can investigate this further by considering the
transverse separation
cutoff ($b_c$) dependence of the form factor.  This can give an idea about the
integration regions important for the calculation. We show the $b_c$ dependence
in fig. \ref{pion3} as originally discussed in Ref. \cite{LS}. Based on
this plot Li and
Sterman argue that roughly 50\% of the contribution can regarded
as perturbative, since it is obtained
from the region where $\alpha_s/\pi < 0.7$.
The observation also implies that higher order contributions in $\alpha_s$
are not
negligible, and the leading order predictions for the normalization of the
form factor cannot be regarded as accurate. The next to leading order
calculation \cite{passek} of the pion form factor also leads to the same
conclusion.

\medskip

We are left with the following interesting situation: Although the basic
Brodsky-LePage
factorization is correct, one may need to go to higher orders in $\alpha_s$
in order to obtain an accurate prediction for the form factor normalizations.
However the predicted $Q^2$ dependence appears
to be quite robust, and independent of the theoretical uncertainties
such as the choice of distribution amplitude.

\section{THE PROTON}

The improved factorization has also been applied to the proton Dirac form
factor $F^p_1(Q^2)$\cite{L}.
The calculation is considerably more complicated compared to the pion.
Here also it is necessary to use distribution amplitudes which peak close to
the end points in order to obtain the experimental normalization
of the form factor. In contrast to pion, there is no
natural choice for the infrared cutoff $\omega$ in the Sudakov
exponent, due to the presence of three quarks and resulting three
distances.

The Sudakov resummation of large logarithms in $\cal P$ leads to
\begin{equation}
{\cal P}(x_i,{\bf b}_i,P,\mu)
=\exp\left[-\sum_{l=1}^3 s(x_l,w,Q)-3\int_{w}^{\mu}
\frac{d\bar{\mu}}{\bar{\mu}}\gamma_{q}\left(\alpha_s(\bar{\mu})
\right)\right]
\times\phi(x_{i},w)\; ,
\label{8}
\end{equation}
where the quark anomalous dimension $\gamma_{q}(\alpha_s)=-\alpha_{s}/\pi$
in axial gauge governs the RG evolution of $\cal P$.
The function $\phi$ is the standard proton distribution amplitude.
The exponent $s$ is given in Ref. \cite{L}.

In equation \ref{8} we use the same infrared parameter $\omega$ for all
the three $s(x_l,\omega,Q)$ for $l=1,2,3$ as well as in the integrals
over the anomalous dimension. Earlier
Li \cite{L} chose to use different infrared cutoffs
 $b_l$
for each exponent $s$ and for each integral involving $\gamma_q$.
As pointed out in \cite{JKB} this choice does not
does not suppress the soft divergences
from $b_l\to 1/\Lambda$ completely, where $\Lambda$ is the QCD
scale parameter. For example, the divergences from
$b_1\to 1/\Lambda$, which appear in $\phi(x_i,w)$ as $w\to \Lambda$,
survive as $x_1\to 0$, since
$s(x_1,b_1,Q)$ vanishes and $s(x_2,b_2,Q)$ and $s(x_3,b_3,Q)$ remain finite.
On the other hand, $w$ should play the role of the factorization scale,
above which QCD corrections give the perturbative evolution of the wave
function ${\cal P}$ in Eq.~(\ref{8}), and below which QCD corrections are
absorbed into the initial condition $\phi$. It is then not
reasonable to choose the cutoffs $b_l$ for the Sudakov resummation
different from $w$.

A modified choice of the cutoffs,
 $w=1/b_{max}$, $b_{max}
=\max(b_l)$, $l=1,2,3$,
was proposed in Ref. \cite{JKB}.
This choice was found to suppress
the soft enhancements, and the form factor was found
to saturate as $b_c \rightarrow 1/\Lambda$.
The authors also included a model non-perturbative
soft wave function in the calculation.
Unfortunately, it turned out that the normalization of the
resulting $Q^4F_1$ was
found to be
less than half of that of the data for all the distribution amplitudes
explored \cite{JKB}. Bolz et al \cite{JKB} then concluded that pQCD
is unable to fit the experimental form factor.

Kundu et al \cite{KLSJ} reexamined the situation.  The group argued that the
 form factor normalization is sensitive to the precise
choice of the infrared cutoff $w$. They used $w=c/b_{max}$, $b_{max}
=\max(b_l)$, $l=1,2,3$, as the infrared cutoff in the Sudakov
exponent,
instead of the one used by Bolz et al \cite{JKB}.
The introduction of  parameter $c$ is
natural from the viewpoint of the resummation, since the scale $cw$,
with $c$ of order unity, is as appropriate as $w$ \cite{BS}.
Kundu et al \cite{KLSJ} find that the calculation is in good agreement
with data
using the King-Sachrajda (KS)
\cite{KS} distribution amplitude and setting $c=1.14$.

The choice $c=1.14$ can also be motivated physically
by considering the proton as a quark-diquark type configuration.
The diquark constituents are the two quarks closest to
each other in the transverse plane. Let $d_{\rm typ}$ be the distance
between the
center of mass of the diquark and the remaining third quark. Then
the infrared cutoff scale $\omega$ can be taken to
be $1/d_{\rm typ}$.
We choose $c$ such that
for a large number of randomly chosen triangles,
 we get for the average
$\langle d_{typ}/b_{max} \rangle=1/c$.
Defining $c$ in such a way, gives
$c\approx1.14$.

\begin{figure} [t,b]
\hbox{\hspace{6em}
 \hbox{\psfig{figure=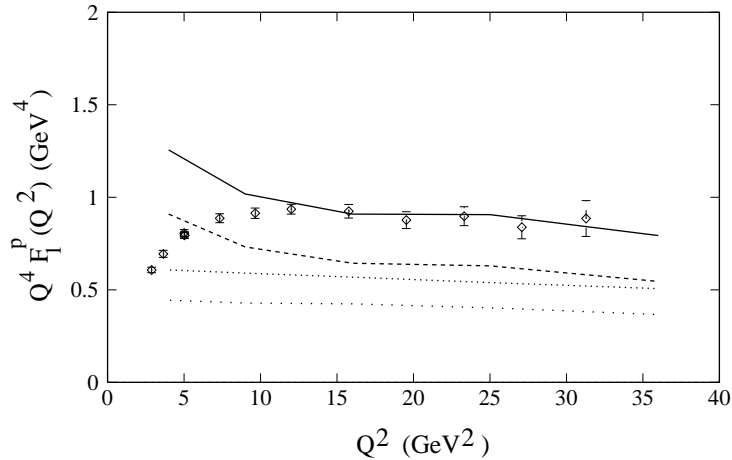,height=6cm}}}
  \caption {\em Dependence of $Q^4F_1^p$ on $Q^2$ using the
    KS distribution amplitude ($c=1.14$, solid line; $c=1$, dense-dot line)
     and the CZ distribution amplitude ($c=1.14$, dashed line;
               $c=1$, dotted line).
    The experimental data with error bars, taken from Ref. 17,
 are also shown.}
  \label{proton1}
\end{figure}

\begin{figure} [t,b]
\hbox{\hspace{6em}
 \hbox{\psfig{figure=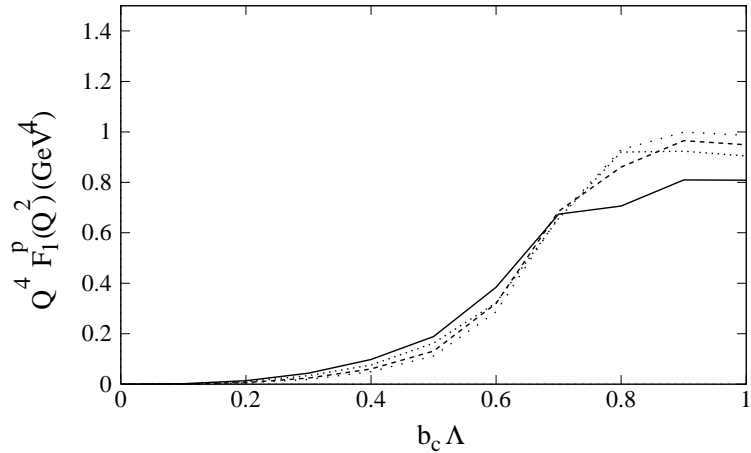,height=6cm}}}
  \caption {\em Dependence of $Q^{4}F_{1}^{p}$ on the cutoff $b_{c}$
    with the KS distribution amplitude for $Q^2=12$ GeV$^2$ (dotted line),
    $Q^2=16$ GeV$^2$ (dashed line), $Q^2=25$
    GeV$^2$ (dense-dot line), and $Q^2=36$ GeV$^2$ (solid line).
	The QCD scale parameter $\Lambda$ was taken to be $0.2$
GeV for this calculation.
}
  \label{proton2}
\end{figure}

\begin{figure} [t,b]
\hbox{\hspace{6em}
 \hbox{\psfig{figure=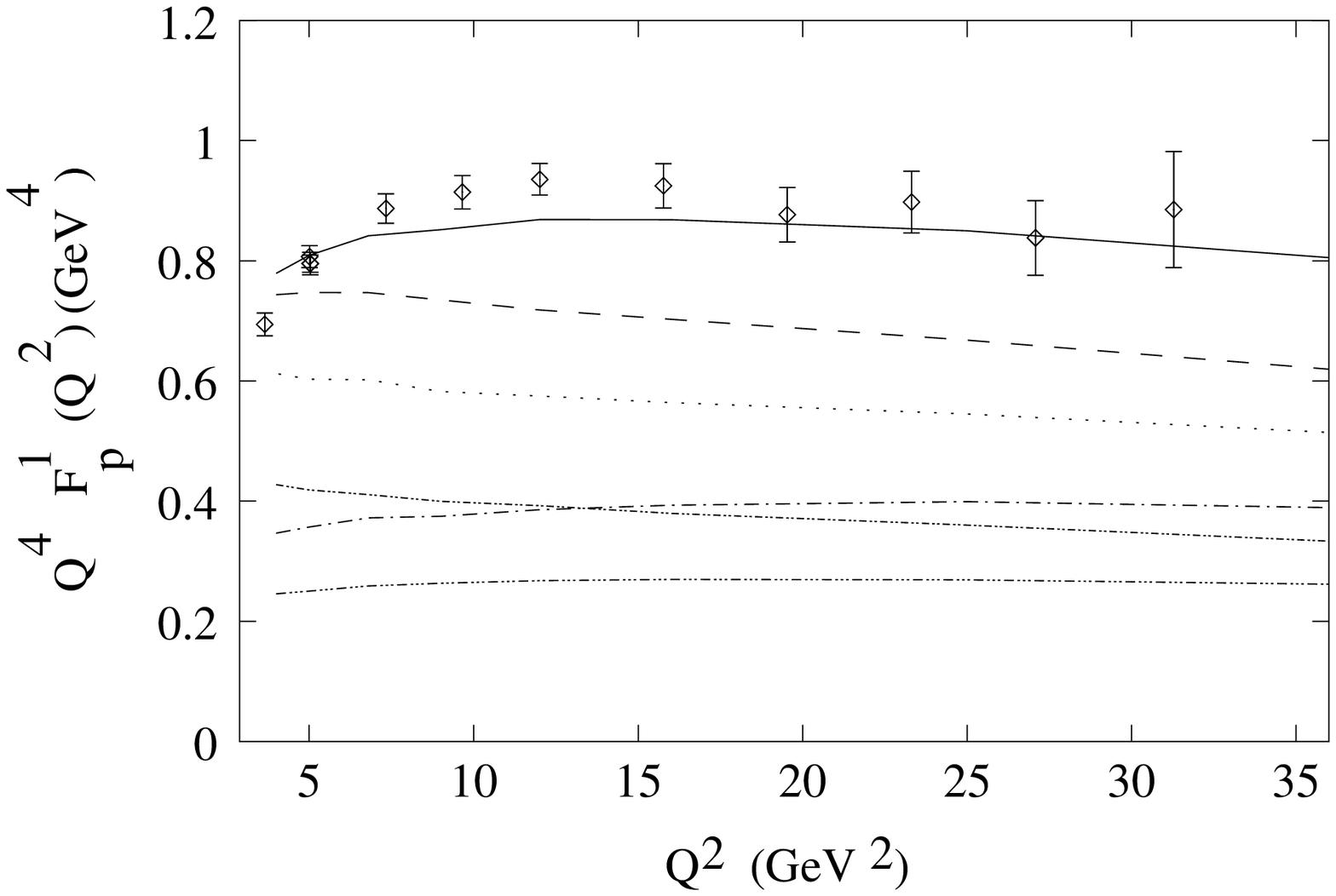,height=6cm}}}
  \caption {\em $Q^2$ dependence of $Q^4F_1^p$ including a soft model
	wave function with $<k_T^2> = 0.271$ GeV$^2$.
The four upper curves at $Q^2 = 35$ GeV$^2$ use the KS
distribution amplitude with the infrared parameter
$c=1.4$ (solid curve), $c=1.2$ (dashed
curve) and $c=1$ (dashed-dotted curve). The dotted curve shows the result
without including model soft wave function with $c=1$. The two lower curves
at $Q^2=35$ GeV$^2$ use CZ distribution amplitude
with $c=1$. The lowest curve includes the
soft model wave function, whereas the upper curve does not. The
    experimental data  with error bars  [17] are also shown.}
  \label{proton3}
\end{figure}

The results of the calculation using KS \cite{KS}
and CZ \cite{CZ} distribution amplitudes and $c=1$ and 1.14
are shown in fig. \ref{proton1}.
It is found that with $c=1.14$ the KS distribution amplitude gives good
agreement with data. The $b_c$ dependence of the form factor is shown in
fig. \ref{proton2}, which shows saturation at about
$b_c = 0.8/\Lambda$.
The result after including
a model nonperturbative soft wave function
are displayed in fig. \ref{proton3}. Again we find that choosing
$c$ of order unity gives pQCD calculations in agreement with data.
For all choices of the distribution
amplitude and parameter $c$, independent of whether the model soft
wave function is included or not, the $Q^2$ dependence of the form factor is
in good agreement with data. An analogous situation was found for pion
form factor.

\medskip

The natural agreement of $Q^{2}$ dependence of the pQCD calculations
is in contrast to data fits
obtained using soft model \cite{Rady,Diehl}. In such models the
$Q^2$ dependence depends on the details of the
model wave function. Soft model predictions at high momentum have a
tendency to fall
more strongly than experimental data. As also noted for the pion,
the approximate power law behaviour in  $Q^2$ is
not directly implied by the factorization and is a detailed dynamical
prediction of the calculation. This could have been achieved only
if the intrinsic $k_T$ were negligible in the hard scattering kernel.
Thus the observation of power-law dependence in the data
lends considerable support to the basic
factorization of Brodsky-LePage. Nevertheless, the relatively small
magnitude of
internal momentum
scales and the sensitive dependence of final result on the
infrared scale $w$ indicates that calculation of normalizations using
leading order diagrams is not reliable. While higher order
contributions are propably non-negligible, there is every reason to
believe that the power-law dependence of the calculations is robust.

We have reviewed the current status of pQCD
calculations of hadronic electromagnetic
form factors. We argue that the normalization of the form factor
cannot be predicted reliably by a leading order calculation in $\alpha_s$.
Detailed calculations including the soft $k_T$ dependence, however, support
the basic factorization scheme. One finds the correct asymptotic $Q^2$
evolution
of the form factors for $Q$ as small as 2-3 GeV, independent of the choice
of distribution amplitude and other theoretical uncertainties.
Hence we conclude that agreement of quark counting scaling predictions is
not accidental but
is well supported by detailed dynamical calculations.

PJ and BK thank the staff of ICTP, Trieste, for hospitality during a
visit where this paper was written. This work was supported by
BRNS grant No. DAE/PHY/96152,
the National Science Council of the
Republic of China Grant No. NSC-87-2112-M-006-018,
 the Crafoord Foundation,
the Helge Ax:son Johnson Foundation, DOE Grant number
DE-FGO2-98ER41079,
the KU General Research Fund and NSF-K*STAR
Program under the Kansas Institute for Theoretical and Computational
Science.

\end{document}